\documentclass[seceq]{ptptex}
\usepackage{graphicx}
\usepackage{amsmath,amsthm,amssymb}
\notypesetlogo                       
\newcommand{\PSfig}[2]{\includegraphics[width=#1]{#2}}
\newcommand{\Comments}[1]{}
\usepackage[normalem]{ulem}  
\usepackage[dvips]{color} 

\renewcommand\sout{\bgroup \color{red} \ULdepth=-.5ex \ULset}

\renewcommand{\sout}[1]{}
\newcommand{\vp}{{\bf p}}
\newcommand{\vq}{{\bf q}}
\newcommand{\vk}{{\bf k}}
\newcommand{\Ret}{{\rm R}}

\title{%
Entropy Production in Gluodynamics \\
in temporal axial gauge in 2+1 dimensions}
\author{
Akihiro \textsc{Nishiyama}
and
Akira \textsc{Ohnishi}
}
\inst{
Yukawa Institute for Theoretical Physics, Kyoto University,
Kyoto 606-8502, Japan
}

\abst{
Entropy production in non-Abelian gauge theory is discussed in the limit
of vanishing classical fields.
Based on the Kadanoff-Baym equation with the leading order self-energy
in the temporal axial gauge,
the kinetic entropy is introduced and is proven to increase
for a given self-energy.
Numerical simulation of the Kadanoff-Baym equation for the non-Abelian
gauge theory is performed in 2+1 dimensions.
Starting from anisotropic distribution in momentum space,
relaxation to isotropic distribution proceeds and the entropy is produced
due to off-shell scattering of massive transverse polarization mode
as expected from the proof of the H-theorem.
}
\begin{document}
\maketitle

\section{Introduction}
In heavy-ion collisions at the Relativistic Heavy Ion Collider (RHIC),
it is plausible from theoretical arguments and experimental evidences
that a new form of matter made of quarks and gluons
is created.\cite{Whitepapers}
This matter referred to as a strongly coupled QGP (sQGP)\cite{Gyulassy:2004zy}
is characterized by its strongly interacting nature;
the shear viscosity is extremely small
and the thermalization time is very short.
Hydrodynamic models successfully describe the radial and elliptic flows
at RHIC,\cite{HK,Hirano:2005xf}
and they require a local thermal equilibrium in the early stage.
Comparison with the experimental results of the elliptic flow
suggests that a thermalization time of the partons
is about 0.6-1 fm$/c$.\cite{HK}
This short thermalization time seems to contradict
the perturbative estimate,\cite{BMSS}
and it is compatible with the formation time of partons.
This means that the Boltzmann equation would have serious problems
in describing the thermalization
of dense systems, such as glasma, formed in the early stage of collisions. 

At lower energies,
hadronic transport models based on the Boltzmann equation 
have been successfully applied to heavy-ion collisions
including nuclear fragment formation,\cite{AMD}
flavor production,\cite{RQMD}
particle spectra,\cite{JAM}
and
collective flows.\cite{Flow}
By comparison,
hadronic transport models fail to describe early thermalization
and predict smaller elliptic flows at RHIC.\cite{RHIC-HadronTransport}
These successes and failures imply the formation of QGP at RHIC,
and may also suggest the limitation of the Boltzmann equation in dense systems.
Boltzmann approaches are based
on the assumption that
the particle interaction is rare and three particle collisions can be ignored,
and the spectral function is narrow enough
for the quasi-particle approximation to be valid.
In dense matter formed at RHIC,
one or both of these conditions are not satisfied.
One of the prescriptions to implement multi-particle interaction
is to apply the classical field
theories.\cite{MV,ALMY,DN,RV2006,KMOS2009}
When the gluon density is so high that gluons are combined to form
finite color glass condensate, a large part of the energy density is stored
in the classical field.
It is generally believed that the early thermalization is realized
from rapid growth of intense classical Yang-Mills fields
triggered by the plasma instabilities.\cite{MV,ALMY,DN,RV2006,Weibel,NielsenO}
The decay of strong gluon field into quarks and gluons needs
quantum field description in nonequilibrium conditions.
In addition, classical field dynamics should not be applied
to high momentum modes even in the dense early stage.
Another aspect of dense systems, the spectral function change,
can be described perturbatively or by using some kind of resummation
in quantum field theories.
Among various ways of resummation,
the $\Phi$-derivable approach would be the most favorable,
since it does not have the secularity problem,
and leads to a correct equilibrium.
Hence we should study the gluon thermalization processes based on
nonequilibrium quantum field theories
such as the Kadanoff-Baym equation based on $\Phi$-derivable approximation.

The Kadanoff-Baym (KB) equation, equivalently the Schwinger-Dyson equation,
is given by Kadanoff and Baym \cite{BK,KB62}
in the reformulation of a functional technique
with Green's functions by Luttinger and Ward \cite{LW}.
It is related to variational technique of an effective action 
using the so-called $\Phi$-derivable approximation,
and is shown to satisfy the conservation laws of a system.\cite{Baym}
In the $\Phi$-derivable approximation,
diagrams contributing to the self-energy are truncated
with some expansion parameter.
This technique is applied to relativistic systems
and reformulated for nonequilibrium many body systems
by using the Schwinger-Keldysh path integral formalism
and 2-particle irreducible (2PI) effective action
technique.\cite{Schwinger,Keldysh,CJT,NS1984,CH}

Nonequilibrium quantum field theoretical approach
is a tool to understand a large variety of contemporary problems 
in high energy particle physics, nuclear physics, astrophysics, cosmology,
as well as condensed matter physics.\cite{BergesReview,BergesSerreau}
In cosmology it is applied to describe the reheating process
due to the decay of inflaton field coupled to the Higgs
at the end stage of inflation.\cite{BS2003,AT2008,Tranberg2008,AST2004}
In condensed matter physics it is applied to Bose-Einstein Condensate.
In this approach both conservation law and gapless Nambu-Goldstone excitation \cite{HP1959}
are respected.\cite{BergesReview,vHK2002-3}
It might be also possible to apply it to heavy ion collisions
to describe equilibration processes of partons.\cite{BBW2004}

With respect to numerical simulation,
Danielewicz first carried out a pioneering work in 1984 \cite{Danielewicz} to 
describe heavy ion collisions with spectral functions
at the non-relativistic energies. 
In the relativistic level,
the equilibration processes are simulated
in $\phi^4$ model in the symmetric phase
in 1+1,\cite{AB,BC} 2+1 \cite{JCG} and 3+1 \cite{AST,LM} dimensions. 
Numerical analyses have been also done in the broken phase
in 1+1 \cite{CDM2003} and 3+1 \cite{AST} dimensions.
Recently simulations have been extended to the system
in an expanding background.\cite{Tranberg2008}
Similarly based on the next-to-leading order self-energy in the systematic
$1/N$ expansion,\cite{AABBS}
simulations have been done both in the symmetric \cite{Berges}
and broken \cite{AST2004} phases in the $O(N)$ model.
All of these analyses show that the thermalization proceeds;
distribution functions lose information on the initial condition
and converge to the Bose-Einstein distribution.
Until now the KB equation for the non-Abelian gauge theories has never been
solved numerically. This is because we have several problems
such as gauge covariance of the equation of motion,\cite{AS2002,CKZ2005}
infrared singularity,
and violation of Ward identity.\cite{KajantieKapusta1985}

In this paper we discuss the gluon thermalization
in the temporal axial gauge (TAG)
on the basis of the KB equation with the 2PI effective action
in the leading order of the coupling expansion.
We expect that the off-shell particle number changing processes such as
$g\leftrightarrow gg$
\footnote{{The process $g\leftrightarrow gg$ contributes in the leading order $O(g^2)$, while $gg\leftrightarrow gg$ and 
$g\leftrightarrow ggg$ are the next-to-leading order $O(g^4)$ contributions of the coupling expansion in the
$\Phi$-derivable approximation. In this paper we concentrate on the leading order effects.}}
 contribute to the entropy production
in heavy ion collisions.
These off-shell effects come from the gluon decay 
due to finite spectral width and memory integral in the KB equation.
A part of $gg\leftrightarrow gg$ processes, $gg\leftrightarrow g\leftrightarrow gg$, 
is included in the present treatment via the sequential off-shell processes of $gg \leftrightarrow  g$ and $g\leftrightarrow gg$,
and contributes to thermalization.
In the Boltzmann equation, $g\leftrightarrow gg$ is prohibited,\cite{XuG2007}
and the on-shell $gg\leftrightarrow ggg$ process is the most efficient
in entropy production. 
It is known that the on-shell dynamics predicts a long equilibration time
$\tau_{\rm eq}=2-3\mathrm{fm}/c$ \cite{BMSS},
and it can not explain the early thermalization of glasma.
Hence we investigate off-shell dynamics by using KB equation
and give whether thermalization occurs or not due to the effects
which have been neglected.
The present entropy production mechanism from the off-shell gluons is
different from the classical plasma instabilities,
such as the Weibel~\cite{Weibel,ALMY,DN,RV2006}
and Nielsen-Olsen~\cite{NielsenO} types.
Our approach is also different from classical statistical approach
on a lattice, which conjectures the significance of
the leading order and next-to-leading order resummed loop self-energy.\cite{BSS}
The aim in this paper is to show that the off-shell processes $g \to gg$
can contribute to equilibration of gluons.
For this purpose,
we introduce the concept of kinetic entropy and the H-theorem,
and perform numerical simulations of the KB equation in gluodynamics.
In order to avoid the problems mentioned above, 
we make some prescriptions and approximations,
such as gauge fixing, infrared cutoff 
and simulation of only the transverse scattering contributions.

This paper is organized as follows.
In Sec.~\ref{Sec:KB},
we write down the KB equation in gluodynamics
for both transverse and longitudinal part of Green's functions.
As the self-energy,
we take account of the leading order (LO) terms in the coupling expansion,
which contain the off-shell $g\leftrightarrow gg$ processes. 
In Sec.~\ref{Sec:H-theorem},
we introduce kinetic entropy, and by use of the gradient expansion
as in Refs.~\citen{IKV4,Kita,Nishiyama,Nishiyama:2010wc}
we show an analytical proof of the H-theorem for the given self-energy
in order to examine whether the above leading order off-shell processes
($g\leftrightarrow gg$) can contribute to equilibration or not. 
The concept of entropy should also be extended to the systems
with strong classical fields,\cite{KMOS2009}
which is out of the scope in this paper.
In Sec.~\ref{Sec:Results},
we give numerical results of the Kadanoff-Baym equations
for the transverse scattering processes with massive mode
in a spatially uniform $2+1$ dimensional spacetime
in order to extract the effects of the off-shell 
particle number changing processes.
We find that the off-shell particle number changing processes
can actually contribute to entropy production and thermalization.
We summarize our work in Sec.~\ref{Sec:Summary}.

\section{Kadanoff-Baym equation for gluons in Temporal Axial Gauge}
\label{Sec:KB}

\begin{figure}[tb]
\begin{center}
\PSfig{6cm}{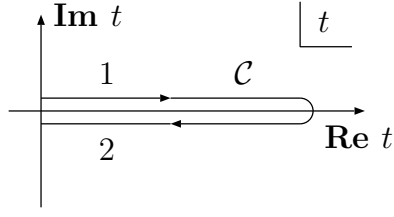}
\end{center}
\caption{
Integral path in the time coordinate.
}
\label{fig:path}
\end{figure}

In this section we write down the Kadanoff-Baym equation for gluons in the temporal axial gauge (TAG), $A_0=0$. 
This section focuses on the practical use of the KB equation.
Let us neglect the classical field $\langle A \rangle =0$ and concentrate on the quantum fluctuations. 
We start with the Lagrangian density
\begin{eqnarray}
{\cal L} = -\frac{1}{4} F^{a \mu \nu} F^a_{\mu \nu}
\ , \quad A_0^a=0
\ ,
\end{eqnarray}
where  $F_{\mu \nu}= \partial_\mu A^a _\nu - \partial_\nu A^a_\mu + gf^{abc}A^b_\mu A^c_\nu $ and 
$f^{abc}$ is a structure constant in non-Abelian gauge theory.
The Greek indices run over the coordinate space $\{0,1,\cdots, d\}$,
and the color indices $\{a, b, \cdots \}$ run over $\{1,\cdots , N^2-1\}$.
 We use a closed time path $\cal C$ along the real time axis,
the path from $t_0$ to $\infty$ and from $\infty$ to $t_0$
as shown in Fig.~\ref{fig:path},
in order to trace nonequilibrium dynamics.\cite{Schwinger, Keldysh}
Then the
2PI effective action with vanishing classical field $(\langle A \rangle =0)$ is written as
\begin{eqnarray}
\Gamma[D]= \frac{i}{2} {\rm Tr}{\ln} (D)^{-1}+ \frac{i}{2} D_0^{-1}D + \frac{1}{2} \Gamma_2[D].
\label{eq:GD}
\end{eqnarray}
Here
$iD_{0\ ij}^{-1}=-(\delta _{ij}\partial_x^2+ \partial_{x, i} \partial_{x, j} )\delta^{ab} \delta_{\cal C}(x-y)$ 
is the free Green's function,
and $D$ $(D^{00}=D^{0i}=D^{i0}=0$; $i,j,\cdots =1,\cdots, d)$ is the full Green's function, 
both of which are defined on the closed 
time path ${\cal C}$. 
The remaining functional $\frac{1}{2}\Gamma_2[D]$ in (\ref{eq:GD}) is 
a sum of the all possible 2PI graphs written with respect to $D$  which remains connected upon cutting
two Green's function lines. %
The stationary relation for the effective action (\ref{eq:GD}),
$\delta \Gamma/\delta D=0$,
gives the Schwinger-Dyson equation for the Green's function $D(x,y)$
\begin{eqnarray}
D^{-1}(x,y) = D_0^{-1}(x,y) -\Pi(x,y)
\label{eq:SDD}
\end{eqnarray}
or equivalently,
\begin{align}
(\delta_{ik}\partial_x^2+\partial_{x,i}\partial_{x,k})D^{ab}_{kj}(x,y)
+i\int_{\cal C}dz \Pi^{ac}_{ik}(x,z)D^{cb}_{kj}(z,y)
=-i\delta^{ab}\delta_{ij}\delta_{\cal C}(x-y)
\end{align}
with the proper self-energy defined as $\Pi=  i \delta \Gamma_2[D]/\delta D$.
The self-energy is expressed as
\begin{eqnarray}
\Pi(x,y)= -i \delta_{\cal C}(x-y)\Pi_{\rm loc}(x)+ \Pi_{\rm nonl}(x,y)
\end{eqnarray}
where the local part $\Pi_{\rm loc}$ contributes to the mass shift while 
the non-local part $\Pi_{\rm nonl}$ induces the mode-coupling between
the different wavenumbers.  

We are now interested in $\{ij\}$ components of
the Green's function and the self-energy,\cite{KajantieKapusta1985}
and shall decompose
the two-point function $D_{ij}^{ab}(x,y)$
and the non-local part of the self-energy $\Pi_{{\rm nonl},ij}^{ab}$
into the statistical and spectral parts,
\begin{align}
F_{ij}^{ab}
=\frac{1}{2} \left[D^{21, ab}_{ij} + D^{12, ab}_{ij} \right]
\ ,&\quad
\rho_{ij}^{ab}
=i \left[D^{21, ab}_{ij}-D^{12, ab}_{ij}\right]
\ ,\\
\Pi_{F, ij}^{ab}(x,y)
= \frac{1}{2} \left[\Pi_{{\rm nonl}, ij}^{21, ab}
+ \Pi_{{\rm nonl}, ij}^{12,ab} \right]
\ ,&\quad
\Pi_{\rho, ij}^{ab}(x,y) 
= i \left[\Pi_{{\rm nonl}, ij}^{21, ab}
- \Pi_{{\rm nonl}, ij}^{12,ab} \right]
\ .
\end{align}
The upper label "1" represents the path from $t_0$ to $\infty$
and "2" represents that from $\infty$ to $t_0$ 
in the closed time path contour $\cal C$. 
The spectral function $\rho(x,y)$ contains the information
on which states are the most realized,
while the statistical function $F(x,y)$ determines
how much a state is occupied. The Schwinger-Dyson
equation (\ref{eq:SDD}) can be equivalently 
rewritten in terms of $F_{ij}^{ab}(x,y)$ and $\rho_{ij}^{ab}(x,y)$ 
as coupled integro-differential equations
\begin{eqnarray}
\left[ (\delta_{ij}\partial_x^2 + \partial_{x,i} \partial_{x,j})
  \delta^{ab} 
+ \Pi_{{\rm loc}, ij}^{ab}(x) \right] 
F_{jk}^{bc}(x,y)
&=& \int_{t_0}^{y^0} dz  \Pi_{F, ij}^{ab}(x,z) \rho^{bc}_{jk}(z,y)
\nonumber\\
&-& \int_{t_0}^{x^0} dz \Pi_{\rho, ij}^{ab}(x,z) F_{jk}^{bc}(z,y), 
\label{eq:KBDF}
\end{eqnarray}
\begin{eqnarray}
\left[ (\delta_{ij}\partial_x^2 + \partial_{x,i} \partial_{x,j})
  \delta^{ab} 
+ \Pi_{{\rm loc}, ij}^{ab}(x) \right] 
\rho_{jk}^{bc}(x,y) 
&=& - \int_{y^0}^{x^0}dz  \Pi_{\rho, ij}^{ab} (x,z) \rho_{jk}^{bc}(z,y)
\label{eq:KBDR}
\end{eqnarray}
where $t_0$ is the initial time.
The set of equations (\ref{eq:KBDF}) and (\ref{eq:KBDR}) is 
the Kadanoff-Baym equation of the gauge theory, which describes 
the time evolution of the fluctuations with $F$ and $\rho$. 
At each time step the spectral function $\rho_{ij}^{ab}(x,y)$ 
must satisfy the conditions following from the commutation relations:
\begin{eqnarray}
\rho_{ij}^{ab}(x,y) |_{x^0\rightarrow y^0} &=& 0,
\nonumber \\
\partial_{x^0}\rho_{ij}^{ab} (x,y) |_{x^0 \rightarrow y^0}
&=& \delta_{ij} \delta^{ab} \delta^d({\bf x-y}), 
\nonumber \\
\partial_{x^0} \partial_{y^0} \rho^{ab}_{ij}(x,y) |_{x^0 \rightarrow y^0} &=& 0.
\end{eqnarray}

In this paper we restrict ourselves
to the spatially homogeneous and color isotropic situation.
From the translational invariance,
we make a Fourier transformation of the Green's functions
and the self-energies with respect to the spatial relative coordinates,
and decompose into
the transverse and longitudinal components 
as\footnote{We adopt the notation $\frac{k_i k_j}{{\bf k}^2} D_L$ instead of 
$\frac{k_i k_j}{{\bf k}^2} \frac{D'_L}{k_0^2} $}
\begin{align}
I^{ab}_{ij}(x^0,y^0;{\bf k}) &=&  \delta^{ab}\left[
 \left(\delta_{ij}- \frac{ k_i k_j}{{\bf k}^2} \right) I_T(x^0,y^0;{\bf k}) 
+\frac{k_ik_j}{{\bf k}^2} I_L(x^0,y^0;{\bf k}) \right],
\end{align}
where $I=D, \Pi, F, \rho, \Pi_F, \Pi_\rho$ or $\Pi_{\rm loc}$.
Then the KB equations of the transverse and longitudinal parts
are simplified in the momentum space as      
\begin{subequations}
\begin{align}
\left[\partial_{x,0}^2 + {\bf k}^2 + \Pi_{{\rm loc}, T}(x^0) \right]
F_T(x^0,y^0;{\bf k})
=& \left\{
   \left[ \Pi_{F,T} \rho_T\right]_{t_0}^{y^0}
  -\left[ \Pi_{\rho,T} F_T\right]_{t_0}^{x^0}\right\}
\label{eq:KB-FT}
\ ,\\
\left[\partial_{x,0}^2 + {\bf k}^2 + \Pi_{{\rm loc}, T}(x^0) \right] 
\rho_T(x^0,y^0;{\bf k}) 
=& -  \left[\Pi_{\rho,T} \rho_T\right]_{y^0}^{x^0}
\label{eq:KB-RT}
\ ,
\end{align}
\end{subequations}
\begin{subequations}
\begin{align}
\left[\partial_{x,0}^2  + \Pi_{{\rm loc}, L}(x^0) \right] F_L(x^0,y^0,{\bf k})
=& 
  \left\{
  \left[\Pi_{F,L} \rho_L\right]_{t_0}^{y^0}
- \left[\Pi_{\rho,L} F_L\right]_{t_0}^{x^0}
  \right\}
\label{eq:KB-FL}
\ ,\\
\left[\partial_{x,0}^2 + \Pi_{{\rm loc}, L}(x^0) \right] 
\rho_L(x^0,y^0;{\bf k}) 
=& - \left[\Pi_{\rho, L} \rho_L\right]_{y^0}^{x^0}
\label{eq:KB-RL}
\ ,
\end{align}
\end{subequations}
where we have introduced a short-hand notation,
\begin{align}
\left[A B\right]_{t_1}^{t_2}
\equiv
\int_{t_1}^{t_2} dz^0 A(x^0,z^0;{\bf k}) B(z^0,y^0;{\bf k})
\ .
\end{align}

\begin{figure}[htbp]
\begin{center}
\PSfig{10cm}{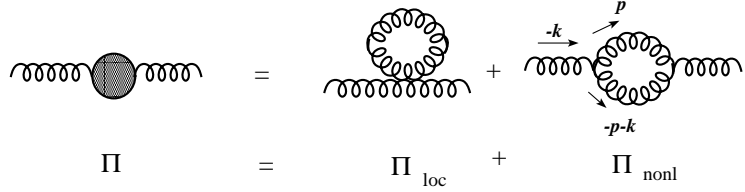}
\end{center}
\caption{Leading order self energy for gluons.}
\label{fig:LOgluon} 
\end{figure}

We approximate the self-energy $\Pi$ with the skeleton
diagrams obtained at the leading order in $g^2$ which 
is depicted in Fig.\ref{fig:LOgluon}. 
The transverse and longitudinal parts are explicitly given
in terms of $D_{T}$ and $D_L$ as
\begin{eqnarray}
\Pi_{{\rm loc},T}(x^0) &=&  g^2 N
\int\frac{d^d p}{(2\pi)^d}
\left[ (d-2) D_T+D_L
 +\frac{\sin^2\theta_{\vp,\vk}}{d-1}(D_T-D_L)
\right]\ ,
 \\
\Pi_{{\rm loc}, L}(x^0) &=&  g^2 N
\int\frac{d^d p}{(2\pi)^d}
\left[
(d-1)D_T+(D_L-D_T)\sin^2\theta_{\vp,\vk}
\right]
\ ,
\end{eqnarray}
where the arguments of $D$ in the integral is $(x^0,x^0;\vp)$.
Similarly the transverse and longitudinal components of
the nonlocal part $\Pi_{\rm nonl}$ are given as
\begin{align}
\Pi_{{\rm nonl},T}(x^0,y^0;-{\bf k})
=&-\frac{g^2 N}{2(d-1)}\int\frac{d^d p}{(2\pi)^d} \sum_{\alpha,\beta}
  P_{\alpha\beta}(\vp,-\vp-\vk) D_\alpha({\bf p}) D_\beta({\bf -p-k}) 
\label{eq:Piii}
\ ,\\
\Pi_{{\rm nonl},L}(x^0,y^0;-{\bf k})
=&-\frac{g^2 N}{2}\int\frac{d^d p}{(2\pi)^d} \sum_{\alpha,\beta}
  Q_{\alpha\beta}(\vp,-\vp-\vk) D_\alpha({\bf p}) D_\beta({\bf -p-k}) 
\label{eq:PiL}
\ ,
\end{align}
where 
the sum runs over $\alpha\beta=TT, LT, TL$ and $LL$.
In the integral,
the temporal variables for $D_\alpha$ and $D_\beta$ are $(x^0,y^0)$.
Here $P_{\alpha\beta}$ and $Q_{\alpha\beta}$ are given as follows
\begin{subequations}
\begin{align}
P_{TT}(\vp,\vq) 
=& P_{TT}^{(1)}(\vp,\vq)+P_{TT}^{(1)}(\vq,-\vp-\vq)+P_{TT}^{(1)}(-\vp-\vq,\vp)
\label{eq:P1}\ ,\\
Q_{TT}(\vp,\vq) 
=& \frac{[(\vp+\vq)\cdot(\vp-\vq)]^2}{(\vp+\vq)^2}\,\left[
		d-2+\frac{(\vp\cdot\vq)^2}{\vp^2\vq^2}
	\right]
\label{eq:Q1}\ ,\\
P_{LL}(\vp,\vq)
=& (\vp+\vq)^2\left[
	1-\frac{(\vp\cdot\vq)^2}{\vp^2\vq^2}
	\right]
\label{eq:P4}\ ,\\
P_{LT}(\vp,\vq) 
=& P_{TL}(\vq,\vp) = Q_{TT}(\vq,-\vp-\vq)
\label{eq:P2}
\ ,\\
Q_{LT}(\vp,\vq)=& Q_{TL}(\vq,\vp) = P_{LL}(\vp,-\vp-\vq)
\label{eq:Q2}\ ,\\
Q_{LL}(\vp,\vq)=&0
\label{eq:Q4}\ ,
\end{align}
\end{subequations}
where $P_{TT}^{(1)}$ is given as
\begin{align}
P_{TT}^{(1)}(\vp,\vq)=&
4(d-1)\left[\vp^2-\frac{(\vp\cdot(\vp+\vq))^2}{(\vp+\vq)^2}\right]
\nonumber\\
  -&\frac43 (\vp+\vq)^2
	\left[1-\frac{(\vp\cdot(\vp+\vq))^2}{\vp^2(\vp+\vq)^2}\right]
	\left[1-\frac{(\vq\cdot(\vp+\vq))^2}{\vq^2(\vp+\vq)^2}\right]
\ .
\end{align}

We find several characteristic features of these coefficients.
First, $P_{\alpha\beta}$ and $Q_{\alpha\beta}$ are all
semi-positive definite for $d\geq 2$,
\begin{align}
P_{\alpha\beta} \geq 0 \ ,\quad Q_{\alpha\beta} \geq 0 \quad (d\geq 2)\ .
\end{align}
Next, we find some symmetric relations.
For example, $P_{TT}$ is symmetric
for any exchange of the variables $\vp, \vq$ and $\vk$,
which satisfy $\vp+\vq+\vk={\bf 0}$,
\begin{align}
P_{TT}(\vp,\vq)=P_{TT}(\vq,\vk)=P_{TT}(\vk,\vp)=P_{TT}(\vq,\vp)
\label{eq:Rsym}
\ .
\end{align}
By using the identity,
\begin{align}
{\bf p}^2- \frac{[\vp\cdot(\vp+\vq)]^2}{(\vp+\vq)^2}
=& 
{\bf q}^2- \frac{[\vq\cdot(\vp+\vq)]^2}{(\vp+\vq)^2}
\label{eq:iden}
\ ,
\end{align}
we find that $P_{TT}^{(1)}$ is symmetric under the exchange of $\vp$ and $\vq$,
and the cyclic sum of $P_{TT}^{(1)}$ leads to the cyclic symmetry.
Other symmetric coefficients ($Q_{TT}$ and $P_{LL}$)
are also symmetric under the exchange of two momentum,
\begin{align}
Q_{TT}(\vp,\vq)=Q_{TT}(\vq,\vp)\ ,\quad
P_{LL}(\vp,\vq)=P_{LL}(\vq,\vp)\ .
\end{align}
For mixed coefficients ($P_{LT}, P_{TL}, Q_{LT}, Q_{TL}$)
Eqs.~(\ref{eq:P2}) and (\ref{eq:Q2}) are rewritten
by using these vectors ($\vp,\vq,\vk$) as,
\begin{align}
P_{LT}(\vk,\vp)=P_{TL}(\vp,\vk)=Q_{TT}(\vp,\vq)\ ,\quad
Q_{LT}(\vp,\vq)=Q_{TL}(\vq,\vp)=P_{LL}(\vp,\vk)\ .
\end{align}
These relations are useful to prove the H-theorem as discussed later.

In the limit of $\vk \rightarrow 0$,
we can reproduce the mass shift
derived by using the Schwinger-Dyson equation in thermal equilibrium,\cite{KajantieKapusta1985}
\begin{align}
\Pi_{ii} (x^0,y^0) =& g^2N (d-1)
 \int \frac{d^d p}{(2\pi)^d} \Big[
-i\delta(x^0-y^0)[(d-1)D_T(x^0,y^0;\vp)+D_L(x^0,y^0;\vp)]
\nonumber\\
& - \vp^2  D_T(x^0,y^0;\vp) \left[2 D_T(x^0,y^0;\vp) + D_L(x^0,y^0;\vp)\right]
\Big]
\ ,
\end{align}
where $\Pi_{ii}=(d-1)\Pi_T+\Pi_L$.
This result is consistent with the hard thermal loop (HTL)
approximation.\cite{KajantieKapusta1985,BI}
As already discussed in Ref.~\citen{KajantieKapusta1985},
the Ward-Takahashi identity is not satisfied
in the KB equation with 2PI effective action,
because diagrams with different orders in $g^2$ are resummed.

\section{Proof of the H-theorem for the non-Abelian gauge theory}
\label{Sec:H-theorem}
In this section we give the expression
of the kinetic entropy in terms of the two-point Green's 
functions $D(x,y)$ for the non-Abelian gauge theory. 
We shall investigate the case with vanishing classical fields
$\langle A\rangle =0$ and color isotropic Green's functions
$D^{ab}(x,y)=\delta^{ab}D(x,y)$ in the temporal axial gauge $A^a_0=0$.
Then we find that the kinetic entropy satisfies the H-theorem.
To be generic, we here consider non-uniform system.
In the leading order of the gradient expansion,
we can still utilize the same coefficients
$P_{\alpha\beta}$ and $Q_{\alpha\beta}$
as those shown for the uniform system in the previous section,
since the modification of these coefficients are in the higher order
in the gradient expansion as discussed later.

The KB equations for the transverse and longitudinal components,
Eqs.~(\ref{eq:KB-FT})-(\ref{eq:KB-RL}),
have a similar form to that for the scalar field theory,
then the kinetic entropy current would have a similar form to that
derived in the scalar theory.\cite{IKV4,Kita,Nishiyama,Nishiyama:2010wc,BIR1999,BIR2001,Riedel1968,VB1998,CP1975}
We start from a conjecture
 that the kinetic entropy current
 per color in the gauge theory with temporal axial gauge is given by
\begin{align}
{s}^\mu \equiv & \int \! \frac{d^{d+1}k}{(2\pi)^{d+1}} (d-1) 
\left[\left (k^\mu-\frac{1}{2} \frac{\partial {\rm Re}\ \Pi_{T, \Ret}}{\partial k_\mu} \right) 
\frac{ \rho_T}{i} 
+ \frac{1}{2} \frac{\partial {\rm Re}\ D_{T, \Ret} }{\partial k_\mu} \frac{ \Pi_{ \rho, T} }{i}
 \right] \sigma(f_T)
\nonumber \\
+&\int \! \frac{d^{d+1}k}{(2\pi)^{d+1}} 
\left[\left (k^0 \delta^{\mu 0}-\frac{1}{2} \frac{\partial {\rm Re}\ \Pi_{L, \Ret}}{\partial k_\mu} \right) 
\frac{\rho_L }{i} 
+ \frac{1}{2} \frac{\partial {\rm Re}\ D_{L, \Ret} }{\partial k_\mu}  \frac{\Pi_{ \rho, L} }{i}
 \right] \sigma(f_L)
\label{s3g}
\end{align}
where 
\begin{eqnarray}
\sigma(f) \equiv  (1+f) {\rm log} |1+f| - f {\rm log} |f| \ ,
\end{eqnarray}
$D_{\alpha, \Ret}=i(D_{\alpha}^{11}-D_{\alpha}^{12})$, $\Pi_{\alpha, \Ret}=i(\Pi_{\alpha}^{11}-\Pi_{\alpha}^{12})$ and 
$f_\alpha$ is defined with expressions $D_{\alpha}^{12}=-i\rho_{\alpha} f_{\alpha}$ and $D_{\alpha}^{21}=-i\rho_{\alpha} (1+f_{\alpha})$
 $(\alpha=T,L)$.
We begin with the Schwinger-Dyson equation Eq.~(\ref{eq:SDD}).
In the similar way as in Ref.~\citen{Nishiyama},
we obtain the following relations with respect to the Fourier transformations
$D(X,k)= \int d(x-y)\ e^{ip\cdot (x-y)} D(x,y)$ where $X=\frac{x+y}{2}$ is the "center-of-mass" coordinate.
\begin{eqnarray}
\partial_{\mu} s^{\mu }(X) = 
\frac{1}{2} \int \! \frac{d^{d+1}k}{(2\pi)^{d+1}} (d-1)C_T {\rm log} \frac{D^{12}_T}{ D_T^{21}} +
\frac{1}{2} \int \! \frac{d^{d+1}k}{(2\pi)^{d+1}} C_L {\rm log} \frac{D^{12}_L}{ D_L^{21}}
\label{eq:Hthg}
\end{eqnarray}
In deriving of the above relation we have used gradient expansion and taken the 1st order of the
approximation. (If the singularity in the longitudinal mode $\sim \frac{1}{(k^0)^2}$ appears, 
it cancels in the fraction of logarithmic part.) 
See, for example, Ref.~\citen{Nishiyama}
for the detailed derivation of the entropy current and its divergence,
Eqs. (\ref{s3g}) and (\ref{eq:Hthg}),
from the KB equation in Eq. (\ref{eq:SDD}).
Here the term $C$ represents the collision terms,
\begin{eqnarray}
C_{T} (X,k)  &=& i \left[ \Pi_{T, \rho} (X,k) F_{T }(X,k)  - \Pi_{T, F}(X,k)   \rho_T(X,k)   \right]
\ ,
\nonumber \\
C_{L} (X,k)  &=& i \left[ \Pi_{L, \rho}(X,k)   F_{L }(X,k)  - \Pi_{L, F} (X,k)  \rho_L (X,k)  \right]
\ .
\label{eq:colli}
\end{eqnarray}
The difference from the scalar theory is that the kinetic entropy
is the sum of the transverse and the longitudinal part.

The remaining work is to prove that the R.H.S. of (\ref{eq:Hthg}) is positive definite. 
It is convenient to express $\Pi_T(X,k)$
in the collision term of (\ref{eq:Hthg}),
which is obtained from the Fourier transforming with respect to the relative
coordinate $x^0-y^0$ and is given as
\begin{align}
\Pi_T(X,-k) =& - \frac{g^2N}{2(d-1)} \int \! \frac{d^{d+1}k}{(2\pi)^{d+1}}
\sum_{\alpha,\beta}
P_{\alpha\beta}(\vp,-\vp-\vk)
D_\alpha(X,p) D_\beta(X,-p-k) 
.
\end{align}
In a similar way,
the longitudinal part $\Pi_L(X,k)$ can be expressed by changing 
$D(x^0,y^0,{\bf p}) D(x^0,y^0;{\bf -p-k})$ to $\int \frac{d p^0}{(2\pi)}D(X,p) D(X,-p-k) $ in (\ref{eq:PiL}).
The collision term (\ref{eq:colli}) is in the first order
of the gradient expansion \cite{IKV4},
and the spatial point dependence
of $P_{\alpha\beta}$ and $Q_{\alpha\beta}$
is in the second order.
Thus we can adopt the coefficients
$P_{\alpha\beta}$ and $Q_{\alpha\beta}$
in the uniform system
by neglecting higher order derivative terms.

In the end we can expand the R.H.S. of (\ref{eq:Hthg}) by 
reexpressing by using $D_{T,L}^{12}$ and $D_{T,L}^{21}$:
\begin{align}
\partial_\mu s^{\mu}
=& \frac{g^2N}{4} \int d\Gamma_{pqk} \sum_{\alpha\beta}
\left\{P_{\alpha\beta}(\vp,\vq)
\left[ K^{12}_{\alpha\beta T}(pqk)
      -K^{21}_{\alpha\beta T}(pqk)\right]
\ln \frac{D_T^{12}(k)}{D_T^{21}(k)}
\right.
\nonumber \\
&~~~~
+\left.Q_{\alpha\beta}(\vp,\vq)
\left[ K^{12}_{\alpha\beta L}(pqk)
      -K^{21}_{\alpha\beta L}(pqk)\right]
\ln \frac{D_L^{12}(k)}{D_L^{21}(k)}
\right\}
\label{eq:Hthg2}
\ ,\\
d\Gamma_{pqk}=&
\frac{d^{d+1}p}{(2\pi)^{d+1}}
\frac{d^{d+1}r}{(2\pi)^{d+1}}
\frac{d^{d+1}k}{(2\pi)^{d+1}}
(2\pi)^{d+1} \delta^{d+1} (p+k+q) 
\ ,\\
K^{12,21}_{\alpha\beta\gamma}(pqk)=&
 D_\alpha^{12,21}(X,p)D_\beta^{12,21}(X,q)D_T^{12,21}(X,k)
\ ,
\end{align}  
where we have used the relation $D_{T,L}^{21}(-k)=D_{T,L}^{12}(k)$.
In R.H.S. of (\ref{eq:Hthg2}) we have three types of three gluon
interaction processes, that is 
three transverse (TTT),
two transverse and one longitudinal (TTL)
and one transverse and two longitudinal processes (TLL).
We have no three longitudinal processes which do not exist for 
any dimensions since $Q_{LL}=0$.

For three transverse (TTT) processes,
only those terms with $P_{TT}$ contribute,
then we apply the symmetry relation (\ref{eq:Rsym}),
which tells us that $P_{TT}$ is symmetry under any exchange of $\vp,\vq,\vk$.
Thus we can omit the variables in $P_{TT}$ in the integral,
$P_{TT}(\vp,\vq)=P_{TT}(\vq,\vk)=\cdots\equiv P_{TT}(pqk)$.
The exchange of variables $k\leftrightarrow p$ and $k\leftrightarrow r$
and averaging of the (TTT) part in Eq.~(\ref{eq:Hthg2}) read:
\begin{eqnarray}
({\rm TTT})
&=& \frac{g^2 N}{12}\int d\Gamma_{pqk}
P_{TT}(pqk) \left[K^{12}_{TTT}(pqk)-K^{21}_{TTT}(pqk)\right]
 \ln \frac{K^{12}_{TTT}(kpr)}{K^{21}_{TTT}(kpr)}
\geq 0
\ ,
\nonumber\\
\label{eq:TTT}
\end{eqnarray}
because $P_{TT}\geq 0$ and $(x-y)\ln(x/y)\geq 0$.
Here note that entropy production is zero for three gluons with collinear momenta due to the form of $P_{TT}$.
Equality holds when the following relation is satisfied;
\begin{eqnarray}
\ln \frac{D_T^{21}}{D_T^{12}} = \ln \frac{1+f_T(X,p)}{f_T(X,p)} =  \beta^\mu (X) p_\mu
\end{eqnarray}
where $\beta(X)$ are arbitrary functions for the center of coordinate $X$.
Thus we note that the local Bose-Einstein distribution function
is realized in equilibrium,
\begin{eqnarray}
f_T(X,p)= \frac{1}{e^{\beta^\mu(X)  p_\mu}- 1}
\ .
\end{eqnarray}

For (TTL) processes,
we have contributions from $P_{LT}$, $P_{TL}$ and $Q_{TT}$ terms.
By exchanging the momentum
$p\leftrightarrow k$ and $q\leftrightarrow k$
in $P_{LT}$ and $P_{TL}$ terms, respectively,
and by using the relation
$P_{LT}(\vk,\vq)=P_{TL}(\vp,\vk)=Q_{TT}(\vp,\vq)$,
we get
\begin{align}
&({\rm TTL})
= \frac{g^2 N}{4}\int d\Gamma_{pqk}
\left[K^{12}_{TTL}(pqk)-K^{21}_{TTL}(pqk)\right]
\nonumber \\
&\times \Big[ 
  P_{LT}(\vk,\vq) \ln \frac{D_T^{12}(X,p)}{D_T^{21}(X,p)}
+ P_{TL}(\vp,\vk) \ln \frac{D_T^{12}(X,q)}{D_T^{21}(X,q)}
+ Q_{TT}(\vp,\vq) \ln \frac{D_L^{12}(X,k)}{D_L^{21}(X,k)}\Big] 
\nonumber \\
&= \frac{g^2N}{4}\int d\Gamma_{pqk}
Q_{TT}(\vp,\vq)
\left[K^{12}_{TTL}(pqk)-K^{21}_{TTL}(pqk)\right]
\ln \frac{K^{12}_{TTL}(pqk)}{K^{21}_{TTL}(pqk)}
\geq 0
\label{eq:TTL}
\ .
\end{align}

For (TLL) processes,
we have contributions from $P_{LL}$, $Q_{LT}$ and $Q_{TL}$ terms.
In a similar way to the (TTL) processes,
by exchanging the momentum
$q\leftrightarrow k$ and $p\leftrightarrow k$
in $Q_{LT}$ and $Q_{TL}$ terms, respectively,
and by using the relation
$Q_{LT}(\vp,\vk)=Q_{TL}(\vk,\vq)=P_{LL}(\vp,\vq)$,
we get
\begin{align}
&({\rm TLL})
= \frac{g^2N}{4}\int d\Gamma_{pqk}
P_{LL}(\vp,\vq)
\left[K^{12}_{LLT}(pqk)-K^{21}_{LLT}(pqk)\right]
\ln \frac{K^{12}_{LLT}(pqk)}{K^{21}_{LLT}(pqk)}
\geq 0
\label{eq:TLL}
\ .
\end{align}

As a result the divergence of the entropy current for the non-Abelian gauge theory in the temporal axial gauge
is found to be positive definite:
\begin{eqnarray}
\partial_\mu s^\mu = ({\rm TTT}:(\ref{eq:TTT})) + ({\rm TTL}:(\ref{eq:TTL})) + ({\rm TLL }:(\ref{eq:TLL})) \geq 0.
\end{eqnarray}
Therefore the off-shell processes of gluons are shown to increase the entropy
and to help the thermalization in the leading order gradient expanded
KB equation in TAG.

There are several comments on the entropy production from off-shell
processes in order.
First, the gluon off-shell processes include
$0 \leftrightarrow 3$ and $1 \leftrightarrow 2$ processes.
These processes are included in KB equation which conserves energy-momentum.
It is seen by taking quasiparticle approximation (zero spectral width)
and remaining (keeping?) memory integral as shown in Ref.~\citen{BergesReview}. 
These off-shell processes contribute to the dynamics except
for the very late time regions where the distribution function change
is sufficiently moderate.
Secondly, in the leading order of the gradient expansion,
$0 \leftrightarrow 3$ processes are forbidden by the energy-momentum 
conservation, while they have finite contributions when the memory integral
is evaluated explicitly without the gradient expansion.\cite{BergesReview}
Thirdly, the $1 \leftrightarrow 2$ processes correspond
to Landau damping diagrammatically and can be treated as the Vlasov term.
In this paper we treat them as collision terms
without taking the HTL approximation,
and they are shown to contribute to entropy production.
One may wonder why does the Vlasov dynamics generate entropy.
Entropy production in quantum systems always requires coarse graining.
The $1 \leftrightarrow 2$ processes induce the non-linearity in the equation of motion
for spectral and statistical functions. As a result, the distribution function 
becomes complex in phase space, and a chaotic behavior will emerge.
The gradient expansion corresponds to the coarse graining in time,
then it induces the information loss of the distribution function,
namely the entropy production.

To complete the proof of the H-theorem,
we need to solve the gauge dependence problem.
Around the equilibrium, the controlled gauge dependence was proven
in Refs.~\citen{AS2002} and \citen{CKZ2005}.
The controlled gauge dependence means that
the gauge dependence always becomes higher order
$\delta \Gamma \sim O(g^{2L})$ in the coupling expansion
under the gauge transformation
of the truncated effective action (thermodynamic potential)
$\Gamma \sim O(g^{2L-2})$, where $L$ is the number of loops.
Thus entropy density around the thermal equilibrium~\cite{Nishiyama:2010wc}
derived by differentiating $\Gamma$ (thermodynamic potential) 
with temperature has controlled gauge dependence.
Unfortunately there is no proof of the controlled gauge dependence
under nonequilibrium conditions,
and its proof is out of the scope of this paper.

\section{Numerical simulation of the transverse mode}
\label{Sec:Results}

While there are several problems in numerical simulations
as mentioned in the Introduction,
it is important to investigate the KB dynamics of non-Abelian gauge theories
numerically.
Here we perform the numerical simulation of the transverse mode
of the gauge theory in the temporal axial gauge in 2+1 dimensions,
as a first step of non-equilibrium quantum field theoretical investigation
of glasma.
We consider a spatially homogeneous system and solve the dynamical
equations numerically.
In order to trace the time evolution of statistical and spectral functions
in the Kadanoff-Baym equation,
at first we shall use a normal lattice discretization
with respect to momentum space.\cite{IMGM}
We discretize the spatial length $L$ into $2N_s$ grid points
with the spatial lattice spacing of $a_s$.
Here ${\bf k}^2$ is discretized in the following,
\begin{eqnarray}
{\bf k}^2
\rightarrow \sum_{i=1}^2 \frac{4}{a_s^2} \sin^2\left(\frac{a_s k_{n_i}}{2} \right)
\end{eqnarray}
where the momenta ${\bf k_n}$ are discretized according
to ${\bf k_n} = \frac{2\pi{\bf n}}{L}$ with
$n_i(i=1,2)\in \{-N_s,\cdots , N_s\}$.
By taking ${\bf k}^2$ as the above way we can remove most of the lattice artifacts. \cite{IMGM}
We take the following replacement in integrating in the momentum space
\begin{eqnarray}
\int \!\! \frac{d^2 p}{(2 \pi)^2} \rightarrow \frac{1}{S} \Sigma_{n\in \{ -N_s,\cdots , N_s\}^2}
\ ,
\end{eqnarray}
where $S=L^2$ is the spatial area.
The mesh size $N_s$ is chosen to be large enough to extract
the momentum distribution and to satisfy the conversions of the solutions
in each momentum mode.
In the numerical analysis we have used a space lattice with
$ma_s=0.3,\ N_s=30, \ a_t/a_s=0.1, \ N_t=400 $
where $m$ is the mass explained in the next paragraph,
$a_t$ is a temporal step size
and $N_t$ is the number of times
for which we keep the propagator in memory
in order to compute the memory integrals\footnote{We have checked the stability of our numerical
simulations by changing these parameters.}.
Because of the exponential damping of unequal-time Green's functions,
we can introduce the cut-off in the memory integral.

We adopt some simplifications in actual numerical simulations.
We are now interested in the dynamics
with nonlocal self-energy $\Pi_{\rm nonl}$.
Hence in our simulation  we replace the tadpole mass shift
by constant mass $m^2$,
which simulates the thermal mass,
\begin{eqnarray}
\Pi_{{\rm loc}, T} \rightarrow m^2 - {\rm Re} \Pi_{T,{\rm R}} .
\end{eqnarray}
Then we make Green's functions and space-time discretization
dimensionless by $m$.
The retarded transverse self-energy is given as
\begin{align}
{\rm Re} \ \Pi_{T,{\rm R}; vac} (k^0,{\bf k})
=& -\frac{g^2 N}{4} \int \frac{d^dp}{(2\pi)^d}
\frac{ P_{TT}({\bf -p, k+p})
\left(\omega_{\bf p} + \omega_{\bf k+p} \right) }
{ \omega_{\bf p}\omega_{\bf p+k} \left[(\omega_{\bf p}+ \omega_{\bf p+k})^2 - (k^0)^2 \right] }\ , \\
\omega_{\bf p} =&\sqrt{{\bf p}^2+m^2}
\end{align}
which is necessary to remove logarithmic divergence
of the nonlocal self-energy and realize the stability
of the Green's function for momentum cutoff.\cite{JCG}
Furthermore we only consider the dynamics of the transverse modes
and three transverse $(\rm TTT)$ processes.
It might be insufficient to take only them in gauge theory,
but we will see that they play a significant role in the dynamics 
of KB equation. 

Here we should comment on $P_{TT}$ function (\ref{eq:P1}).
This function has ambiguity at three points 
$\vk, \vp$ and  $\vp+\vk= {\bf 0}$ where we must
evaluate the angle between the vector with a zero vector.
For example at the limit $\vk \rightarrow {\bf 0}$ the function
$P_{TT}$ has the possibility to take values
in the range $0 \leq P_{TT} \leq 4\vp^2$.
This means that if $\vk$ approaches ${\bf 0}$
in the parallel direction with $\vp$ or $\vp+\vk$, this function takes zero,
while  if in the orthogonal direction, it takes $4\vp^2$.
In these three points, we set $P_{TT}=0$
(the least entropy production case; Case I),
or
we set $P_{TT}=2\vp^2, 2(\vp+\vk)^2$, and $2\vk^2$
at $\vk, \vp$ and $\vp+\vk={\bf 0}$ (Case II), 
where smoothly changing
function is utilized in connecting
parallel and orthogonal direction around the zero vector.
We trace the dynamics for both of these cases.
In the later discussion, we mainly discuss the results in Case II.
 
\begin{figure}[htbp]
\begin{center}
\PSfig{0.5\hsize,angle=270}{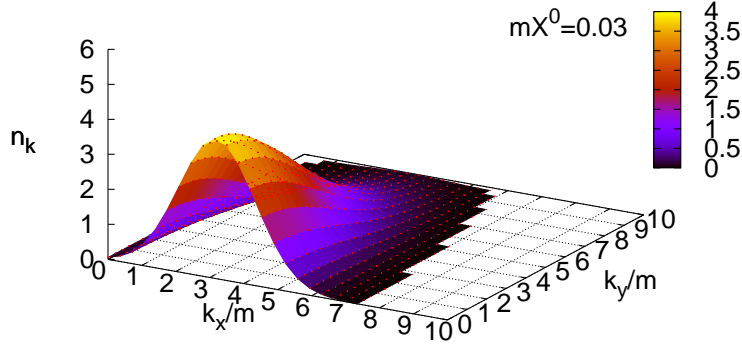}
\end{center}
\caption{Initial distribution function 
$n_{\bf k}^0(mX^0=0.03)$. This function is plotted in $k_x-k_y$ momentum plane.}
\label{fig:ic}
\end{figure}

We use the initial condition for Fourier transformed transverse
spectral functions based on quantum commutation relations
\begin{eqnarray}
\rho_T(x^0,y^0;{\bf k}) \Big|_{x^0=y^0} =& 0\ ,\nonumber \\
\partial_{x^0} \rho_T(x^0,y^0;{\bf k})\Big|_{x^0=y^0} =& 1\ , \nonumber \\
\partial_{x^0} \partial_{y^0} \rho_T(x^0,y^0;{\bf k})\Big|_{x^0=y^0} =& 0. 
\end{eqnarray}
We adopt an initial condition of statistical functions
which contain the information of number distribution functions as
\begin{eqnarray}
F_T(x^0,y^0;{\bf k})\Big| _{x^0=y^0=t_0}
=& \frac{1}{\omega_\vk^0}(n_\vk^0+1/2)
\nonumber \\
\partial_{x^0}F_T(x^0,y^0;{\bf k})\Big|_{x^0=y^0=t_0} =& 0
\nonumber \\
\partial_{x^0}\partial_{y^0} F_T(x^0,y^0;{\bf k})\Big|_{x^0=y^0=t_0}
= & \omega_\vk^0 (n_\vk^0 + 1/2)
\label{fxypi}
\end{eqnarray}
where $n_{\bf k}^0$ is initial number distribution functions,
and $\omega_\vk^0=\sqrt{\vk^2+m^2}$.
We here adopt the so-called "Tsunami" distribution expressed as
\begin{eqnarray}
n_{\bf k}^0= \frac{1}{\cal N} \exp\left(-\frac{(|k_x|-k_t)^2}{2\sigma_1^2} -\frac{k_y^2}{2\sigma^2_2}\right)
\label{eq:ic}
\end{eqnarray}
with $\sigma_1^2/m^2=10.4\times \left(\frac{2\pi}{mL} \right)^2$,
$\sigma_2^{2}/m^2= 100\times \left(\frac{2\pi}{mL} \right)^2$,
$k_t=9.20\cdot 2\pi /L$ and ${\cal N}=0.25$.
This function has Gaussian peaks around $k_x=k_t$
and $k_y=0$, and the widths are different for $k_x$ and $k_y$ directions,
as shown in Fig. \ref{fig:ic}. 
This might be seen as the toy model for gluons in heavy ion collisions.
In the following we omit the subscript $"T"$ in Green's functions.

Now let us discuss the time evolution of the KB equation
for the transverse modes.
We solve (\ref{eq:KB-FT}) and
(\ref{eq:KB-RT}) with mass $m^2$ and self-energy,
which contains three transverse Green's functions.  Time evolution of
Green's functions is performed by referring to and in a similar way as Ref.~\citen{Berges}.
We follow the evolution of $n_{\bf k}(X^0)$
as a function of $\tilde{\omega}_\vp$ defined as
\begin{align}
n_{\bf k} (X^0) +\frac{1}{2}
=& 
\left.\sqrt{
F(\partial_{x^0}\partial_{y^0}F)
-(\partial_{x^0}F)^2
}\right|_{x^0=y^0=X^0}
\label{eq:np}
\ , \\
\tilde \omega_{\bf k}(X^0)
=& \left.\sqrt{(\partial_{x^0}\partial_{y^0}F)/F}\right|_{x^0=y^0=X^0}
\label{eq:op}
\ .
\end{align}
 starting from
the initial condition Eq. (\ref{eq:ic}). 

\begin{figure}[htbp]
\PSfig{5cm,angle=270}{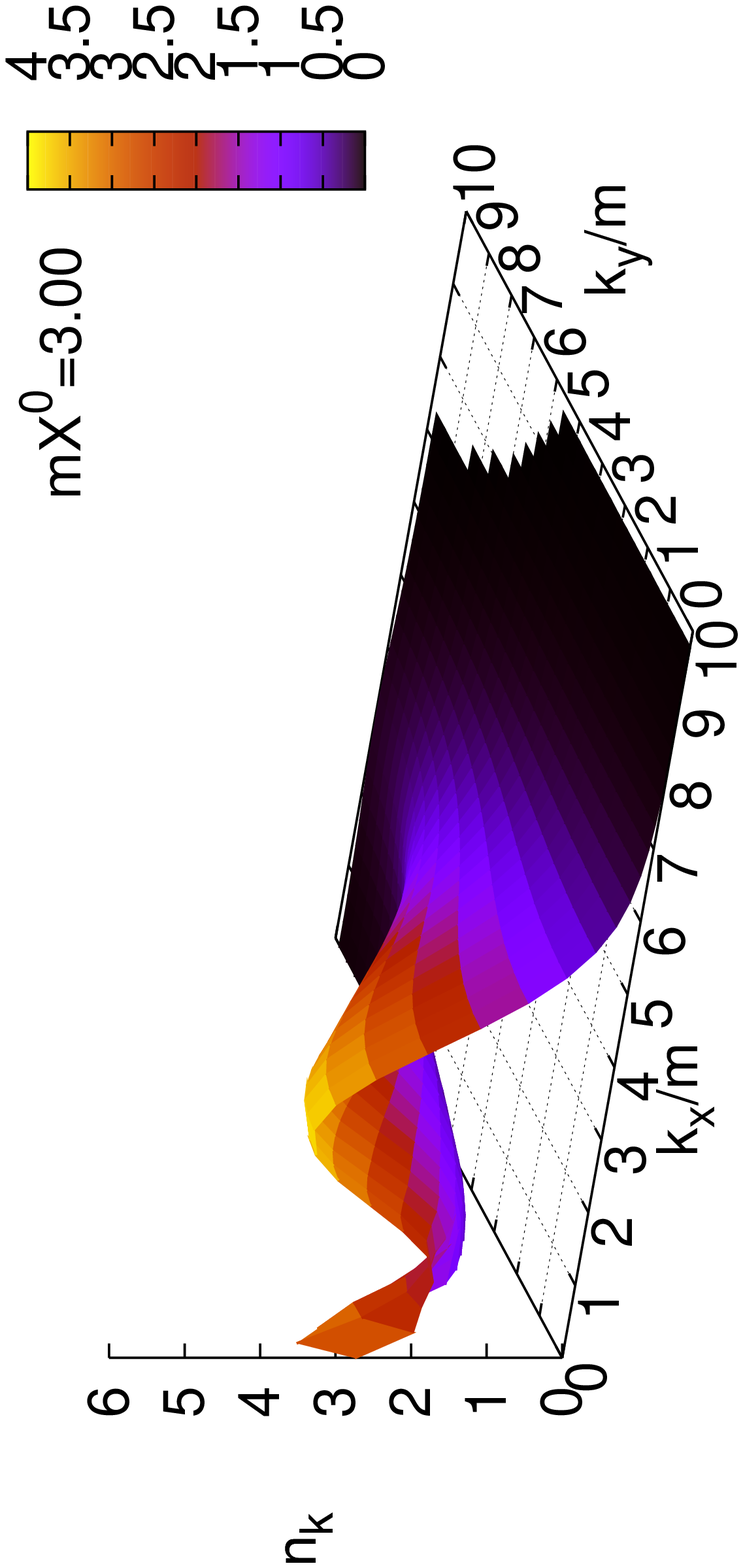}~\PSfig{5cm,angle=270}{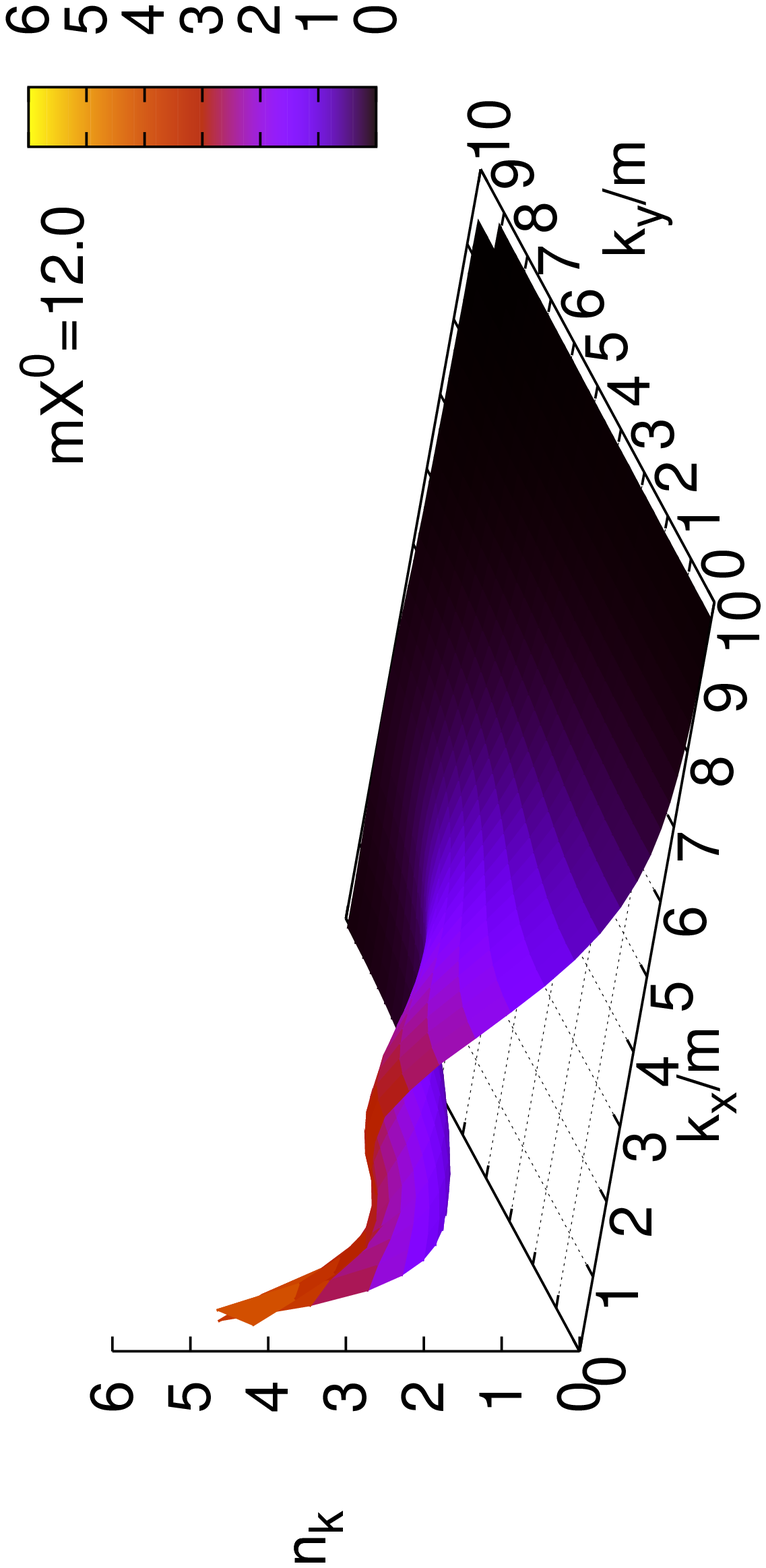}\\
\PSfig{5cm,angle=270}{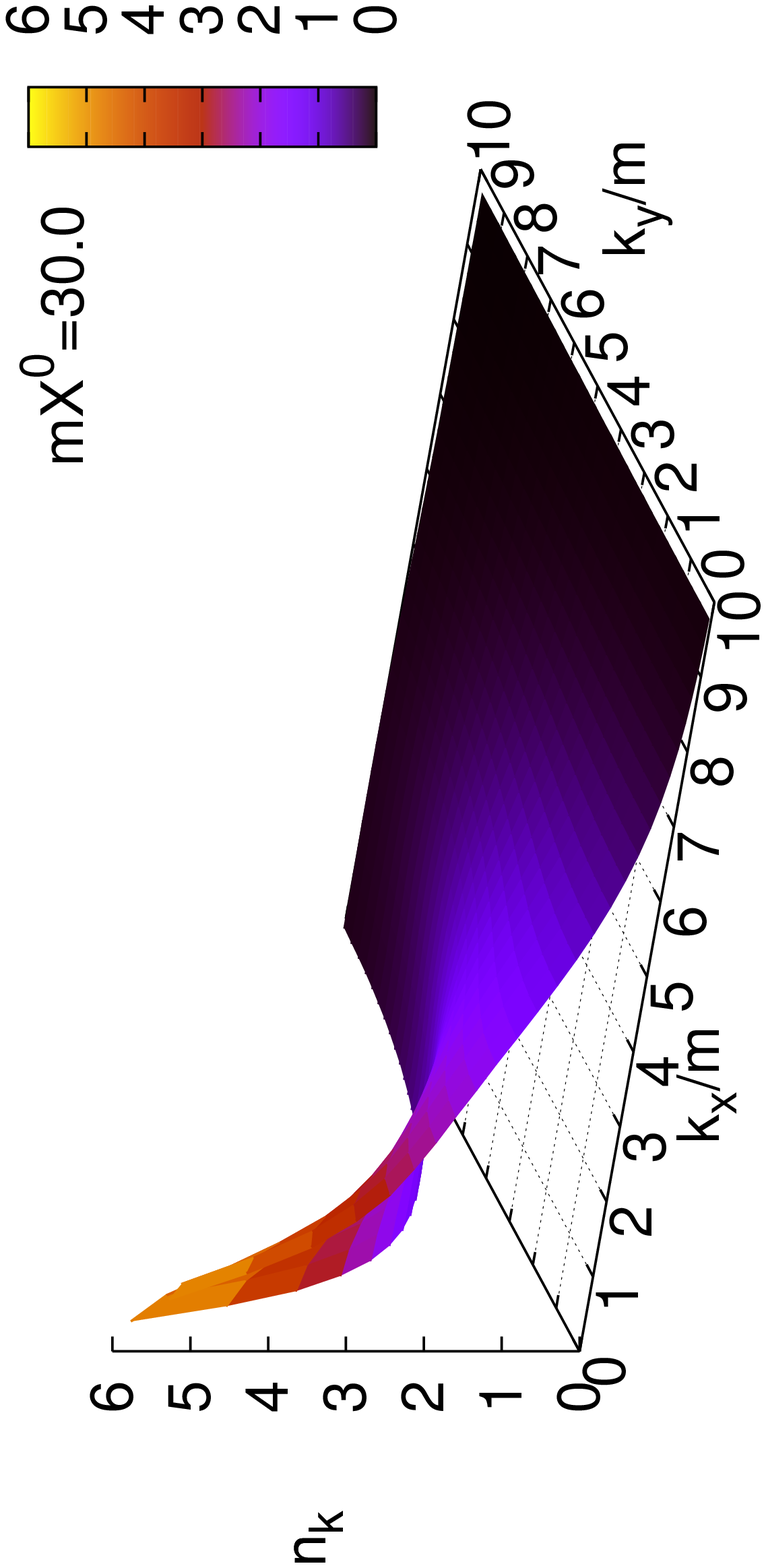}~\PSfig{5cm,angle=270}{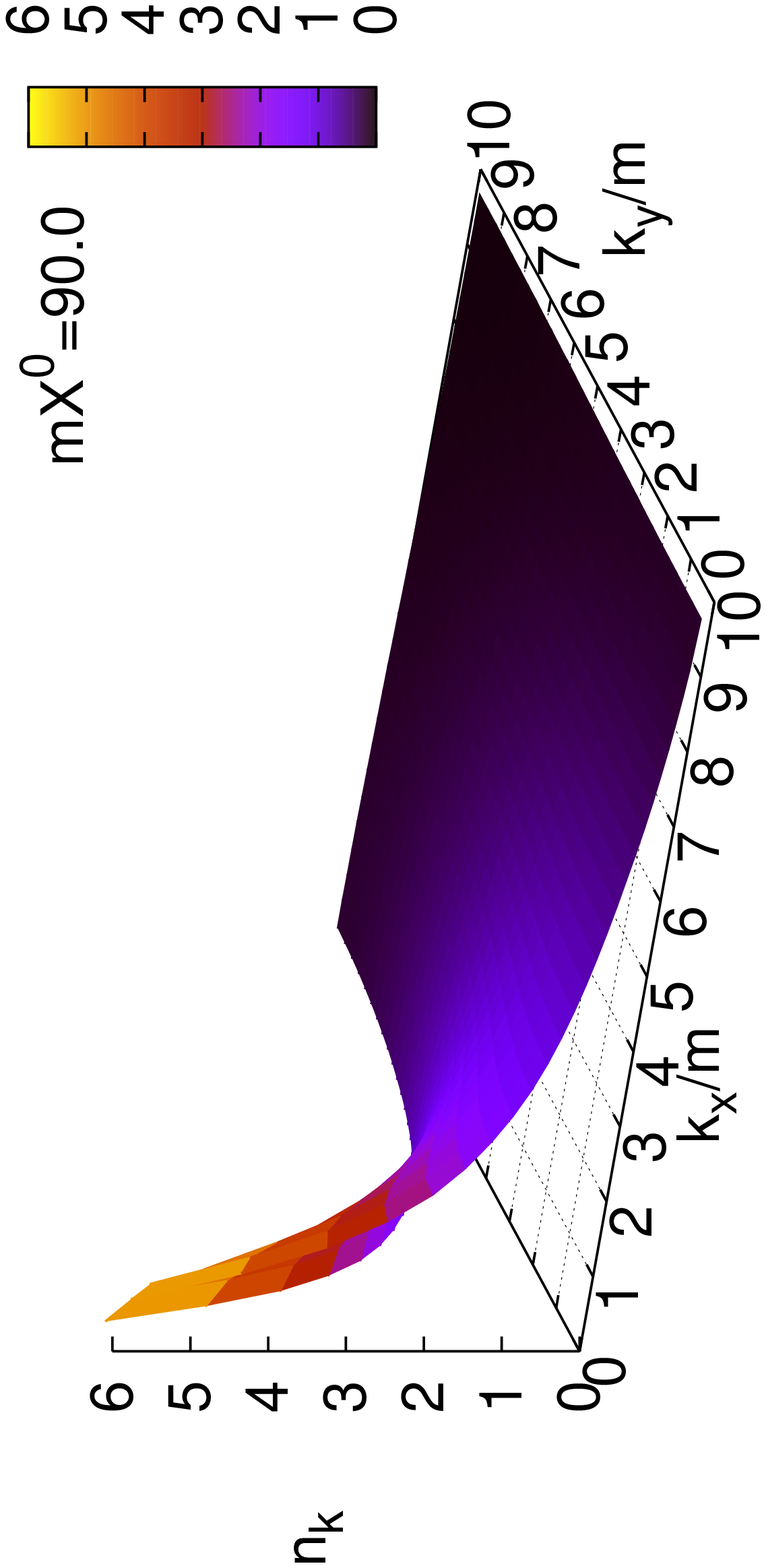}
\caption{Number distribution function 
$n_{\bf k}$ at 
$mX^0=3.0$(upper left), $mX^0=12$ (upper right),
$mX^0=30$(lower left) and $mX^0=90$ (lower right).
The coupling constant and the number of color are
$g^2/m=0.08$ and $N=3$.
}
\label{fig:np}
\end{figure}  

\begin{figure}[htbp]
\begin{center}
\PSfig{5cm,angle=270}{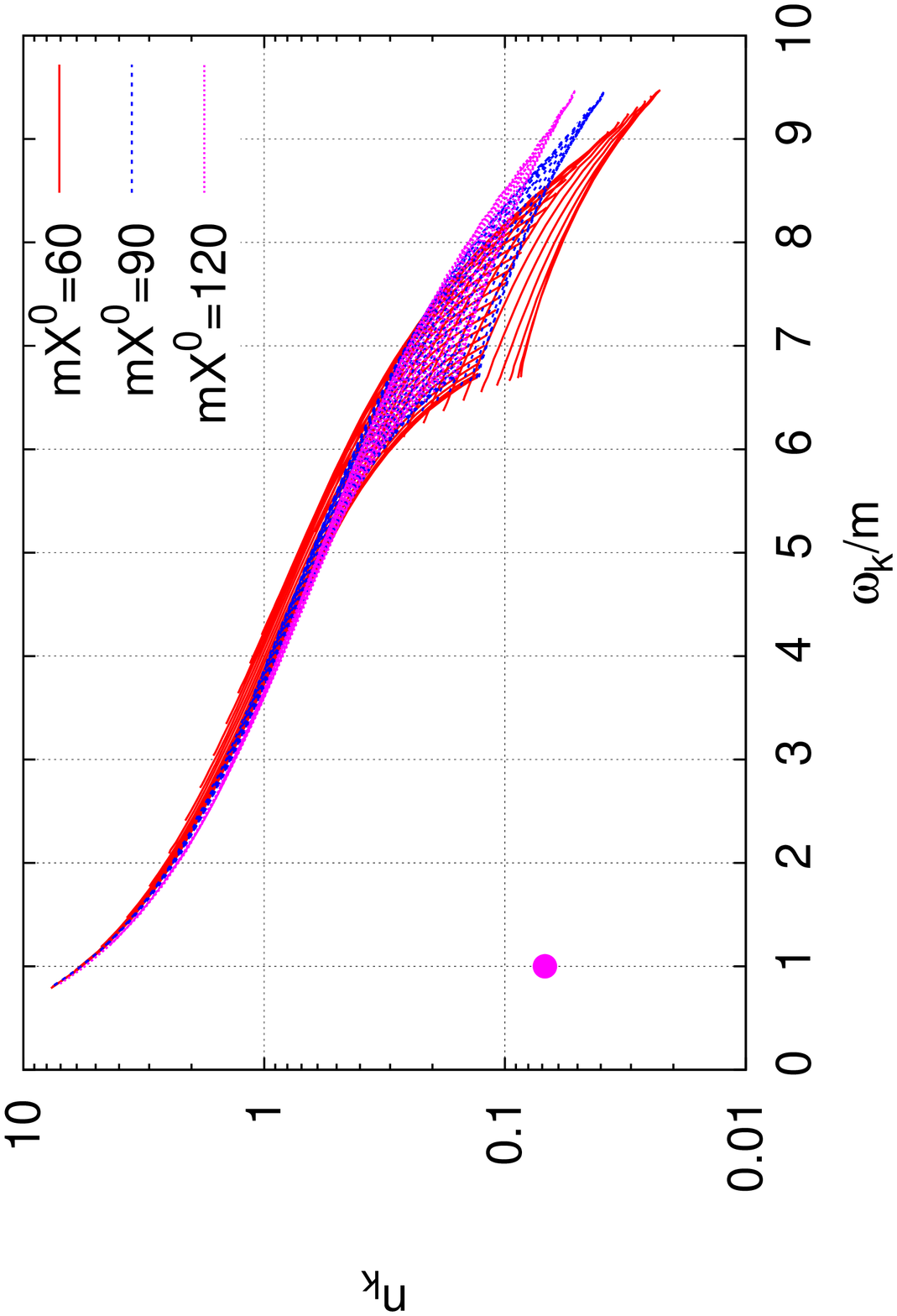}~\PSfig{5cm,angle=270}{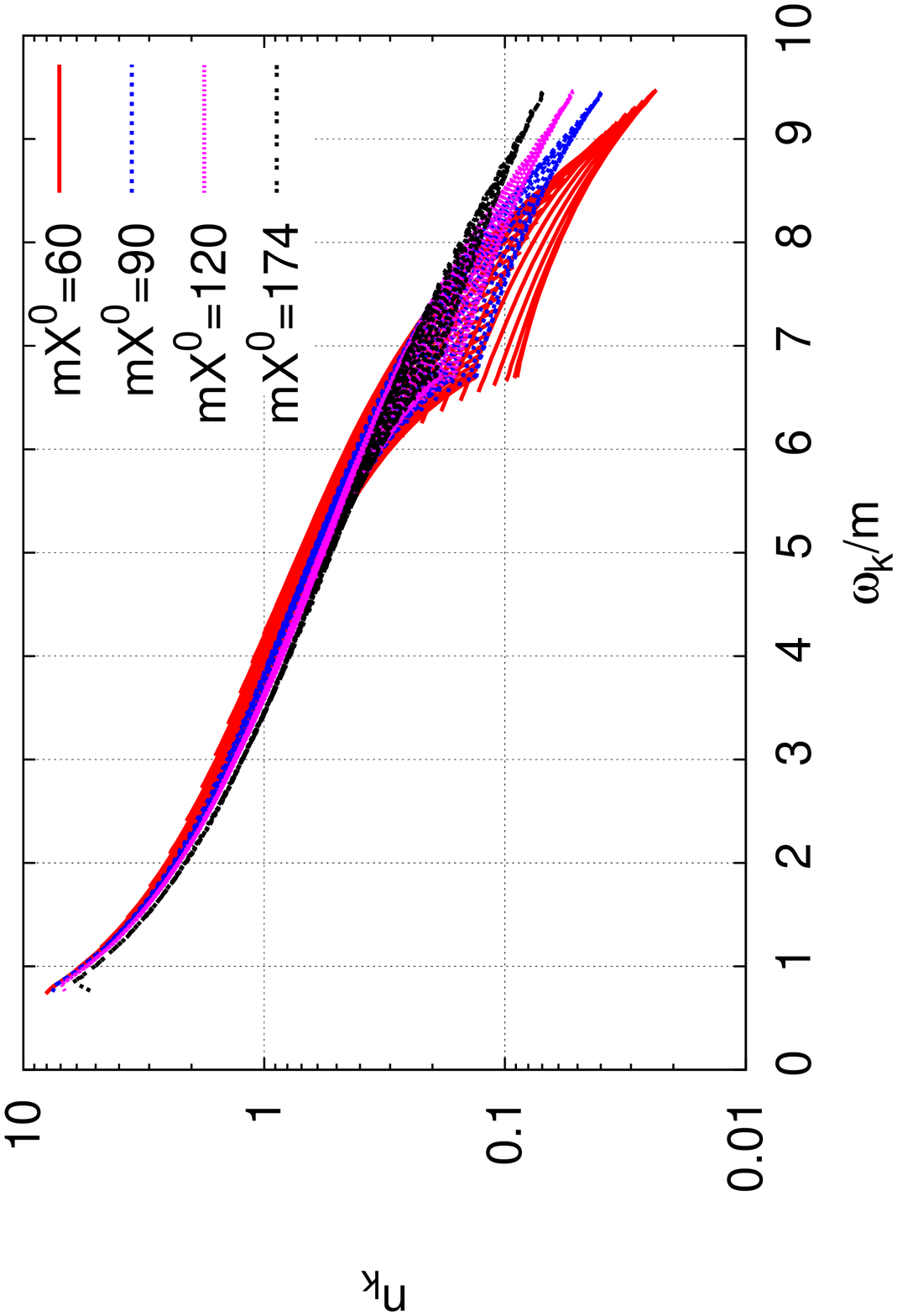}
\end{center}
\caption{Logarithmic plot of number distribution function 
$n_{\bf k}$ at late times.
Left and right panels show the results in Case I and Case II, respectively. 
In Case I, we show the zero momentum mode with the filled circle.}
\label{fig:nplog}
\end{figure}

In Fig.~\ref{fig:np},
we show the time evolution of $n_{\bf k} (X^0)$ at $3.0 \leq mX^0 \leq 90$ with $g^2/m=0.08$ and $N=3$.
In Fig.~\ref{fig:nplog},
we compare the number distribution $n_\vk$ as a function of $\omega_\vk$
in Case I and II.
There are several points to be noted from these distribution.
First, we find that the distribution function converges to the equilibrium
distribution.
At later times,
the number distribution function loses its Gaussian structure
and becomes isotropic in momentum space.
Thus we have confirmed
that off-shell transverse $g\leftrightarrow gg$ processes
can cause kinetic equilibrium.
Secondly, the equilibrium distribution at low energy
is in agreement with the Bose distribution.
At later times,
the temperature and the chemical potential
are found to be
$T \simeq 2.8 (3.2)$ and $\mu \simeq -1.7(-1.4)$
at $mX^0=120 (174)$.
The chemical potential appears to converge to zero as expected
because the number of gluons is not a conserved variable.
Simulations starting from different initial conditions
with the same energy density show the same distributions at late times.
Thirdly,
high momentum modes are found to need longer time to grow up.
For the modes with $\omega>6$,
the distribution function is still different from the Bose distribution
at $mX^0\sim 120$.
Finally, we find that the results are almost the same
for non-zero momentum modes in Cases I and II.
The zero momentum mode of $n_{\bf k}(X^0)$ in Case I
remains the same value since zero momentum mode of the self-energy is always zero,
but it can evolve in Case II and approaches the Bose distribution value at zero momentum.
We can conclude that the ambiguities at $\vk, \vp, \vk+\vp=0$
does not induce any practical problems.

\begin{figure}[htbp]
\begin{center}
\PSfig{8cm,angle=270}{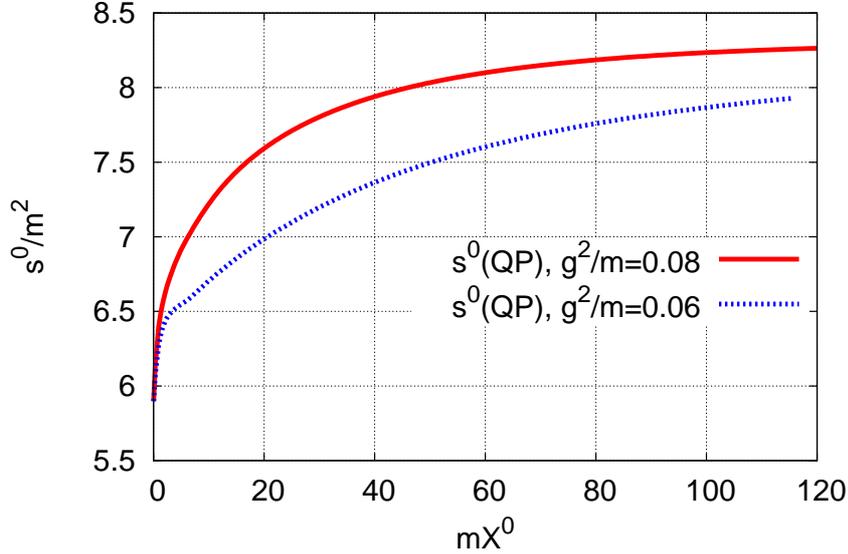}
\end{center}
\caption{Evolution of entropy density in quasiparticle approximation (\ref{eq:entq})
 with $g^2/m=0.08$ (solid) and 0.06 (dotted).}
\label{fig:Ent}
\end{figure}

Figure \ref{fig:Ent} shows the time evolution of the entropy density
$s^0/m^2$ (\ref{eq:entq}) 
with $g^2/m=0.08$ and $0.06$.
Here we employ another entropy density expression,
\begin{eqnarray}
s^0(X) = (d-1) \int \! \! \frac{d^d k}{(2\pi)^d} \left[(1+n_{\bf k}) \ln (1+n_{\bf k}) 
- n_{\bf k} \ln n_{\bf k} \right],
\label{eq:entq}
\end{eqnarray}
which can be a criteria of thermalization,\cite{Nishiyama}
instead of entropy density (\ref{s3g}) to reduce numerical cost.
The entropy density Eq.~(\ref{eq:entq}) is an approximate expression
of Eq.~(\ref{s3g}),
and it can be derived from (\ref{s3g}) in the quasi-particle limit;
by setting $\Sigma_\rho\rightarrow 0$,
replacing $G^{12}=-i\rho f= 2 \pi\delta ((k^0)^{2}- \tilde \omega_{\bf k}^2) (\theta(-k^0)+n_{\bf k})$
and 
$G^{21}= -i \rho (1+f) =
2\pi \delta ((k^0)^{2}- \tilde \omega_{\bf k}^2)(\theta(k^0)+n_{\bf k}) $ 
as in Ref.~\citen{Nishiyama} 
and neglecting longitudinal part.
Entropy density increases monotonically in the whole evolution.
It increases rapidly at early times
when the Gaussian structure collapses,
and it saturates in the later stage
when the kinetic equilibrium is nearly achieved. 
Monotonically increasing evolution is consistent with the H-theorem. 
Thus we confirm again that the leading order self-energy
with off-shell  processes, such as $g\leftrightarrow gg$,
contributes to entropy production and equilibration
in the numerical simulation
without gradient expansion in the temporal axial gauge. 
This entropy production is characteristic in off-shell KB dynamics,
since Boltzmann equation prohibits $g\leftrightarrow gg$.
This property of off-shell effects may help the understanding
of the early thermalization of dense nonequilibrium gluonic system 
in the initial stage of the heavy ion collision.

\begin{figure}[htbp]
\begin{center}
\PSfig{5cm,angle=270}{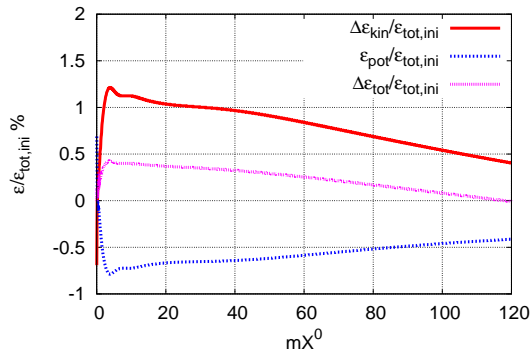}
\end{center}
\caption{Evolution of kinetic (solid), potential (dotted) and total energy (bold solid) in units of m with $g^2/m=0.08$.}
\label{fig:Ene}
\end{figure}

Before closing, we examine the energy conservation in the present
numerical simulation.
The total energy is given by
\begin{eqnarray}
\epsilon_{tot} &=& \epsilon_{kin}+ \epsilon_{pot}
\\
\epsilon_{kin}/L^d &= &  \frac{1}{2} \int \!\! \frac{d^d p}{(2\pi)^d} 
({\bf p}^2+m^2 + \partial_{x^0} \partial_{y^0}   )
 F(x^0,y^0; {\bf p}) \Big | _{x^0=y^0=X^0} 
 \\
 \epsilon_{pot}/L^d &=& \frac{1}{4} \int \!\! \frac{d^dp}{(2\pi)^d} I(X^0, {\bf p}) - \frac{1}{2} \int \!\! \frac{d^dp}{(2\pi)^d} 
{\rm Re} \Pi_{T, {\rm  R}; vac}  F(X^0,X^0; {\bf p}) 
 \nonumber \\
 I(X^0,{\bf p}) &=& \int _0^{X^0} \!\! dt' \left[ \Pi_\rho(X^0,t';{\bf p}) F(t',X^0;{\bf p}) -
 \Pi_F(X^0,t';{\bf p}) \rho (t',X^0;{\bf p}) \right] 
\end{eqnarray}
as shown in Refs.~\citen{Baym,JCG,AST,IKV2}.
Figure \ref{fig:Ene} shows the kinetic, potential and total energy as a function of time $X^0$ for $g^2/m=0.08$ 
measured from their initial values.
We show the results relative to the initial total energy density,
$\epsilon_{tot}/m \sim \epsilon_{kin}/m =18800$ in our lattice spacing.
We find that the growth of the kinetic energy is cancelled by the decrease of
potential energy,
and that the energy error is within 0.5 percent.
This is also true in Case I.

\section{Summary}
\label{Sec:Summary}
In this paper, first we have reviewed the Kadanoff-Baym (KB) dynamics for
the non-Abelian gauge theory
in the temporal axial gauge with the 2PI effective action. 
As an interaction, we have shown the form of leading order self-energy
in KB equation.
In the hard thermal loop (HTL) approximation,
the leading order self-energy contributes
to a mass shift (Debye mass or the thermal mass).
Without the HTL approximation,
nonlocal part of the self-energy 
couples two different momentum modes, and
have a possibility to produce the width of the spectral function
and change the number distribution function during the time evolution. 
By assuming that the system is spatially uniform and isotropic
in the color space, we have decomposed the Fourier transformed Green's 
function into the transverse and longitudinal components,
and expressed the Schwinger-Dyson (SD) equation
in the form of the KB equation, i.e. the coupled evolution equation
of the spectral and statistical functions.
 
We have introduced the kinetic entropy current and given a proof 
of the H-theorem for the leading order self-energy.
From this proof we find that nonlocal part of self-energy can induce
entropy production.
The entropy is produced via the off-shell processes such as $g\leftrightarrow gg$,
and the transverse part of the entropy production, (TTT) in Eq.~(\ref{eq:TTT}),
is shown to stop increasing when the distribution function converges
to the Bose-Einstein distribution.

Finally we have considered 'practical'
time evolution of the transverse modes
of the KB equation numerically,
as a first step of the non-equilibrium quantum field theoretical
simulation of the non-Abelian gauge theory.
We have approximated the local part of the self-energy
by the constant mass term and the retarded local self-energy in vacuum,
and the longitudinal part is ignored.
We have evaluated the time evolution of this
practical KB equation and investigated
the equilibration of distribution functions. 
During the evolution,
the number distribution function loses its initial information
and approaches the Bose distributions.
Off-shell particle number changing processes from the
leading order self-energy
are demonstrated to change the number distribution 
function in numerical simulation without gradient expansion.
We have confirmed the entropy production from the off-shell processes.
It is consistent with the proof of H-theorem.

The Kadanoff-Baym equation has a property that the width of the spectral function
in the $p^0$ space and the memory effects 
allow off-shell collisions,
which vary the momentum distribution of particle and 
induce entropy production.
It is natural to think that these off-shell effects
can contribute more rapid entropy production also in the 3+1 dimensions.
This would be very important
to investigate the entropy production in the heavy ion collisions.
In order to apply the Kadanoff-Baym equation to actual heavy ion collision problems,
it is necessary to include higher order effects in the coupling expansion,
coupling with the color glass condensate (CGC) which appears as the classical field,
and 3+1 dimensional numerical calculations.
Works in these directions are in progress, and will be reported elsewhere.

\section*{ Acknowledgement}

We would like to thank 
Prof. B. M\"uller, Prof. T. Matsui and Dr. H. Fujii
for grateful discussions in both analytical and numerical calculation
of nonequilibrium statistical physics.
Part of this work was done during the Nishinomiya-Yukawa Memorial Workshop
``High Energy Strong Interactions 2010''.
The research in this paper has been supported by JSPS research fellowships for
Young Scientists under the grant number 21$\cdot$6697,
the Global COE Program
"The Next Generation of Physics, Spun from Universality and Emergence",
and the Yukawa International Program for Quark-hadron Sciences (YIPQS).
The numerical calculations were carried out on Altix3700 BX2 at YITP in Kyoto University.
We thank Komaba Nuclear Theory group in University of Tokyo
 for continuing access to their computational facilities.

\end{document}